\documentclass[%
 reprint,
 amsmath,amssymb,
 aps, prl
]{revtex4-2}

\usepackage{graphicx}
\usepackage{graphicx,amsfonts,color,comment,amsmath,hyperref,float}
\usepackage{dcolumn}
\usepackage{bm}
\usepackage{dcolumn}
\usepackage{natbib,aas_macros}
\usepackage{color}
\usepackage{hyperref}
\usepackage{graphicx, epsfig}
\usepackage[dvipsnames]{xcolor}

\begin{document}

\preprint{APS/123-QED}

\title{Ultraheavy Ultrahigh-Energy Cosmic Rays}

\author{B. Theodore Zhang$^{1,7}$}
\email{zhangbing@ihep.ac.cn}
\author{Kohta Murase$^{2,3,4,1}$}
\email{murase@psu.edu}
\author{Nick Ekanger$^5$}
\author{Mukul Bhattacharya$^{2,3,4}$}
\author{Shunsaku Horiuchi$^{5,6,8}$}
\affiliation{$^1$Center for Gravitational Physics and Quantum Information, Yukawa Institute for Theoretical Physics, Kyoto University, Kyoto, Kyoto 606-8502, Japan}
\affiliation{$^2$Department of Physics, The Pennsylvania State University, University Park, PA 16802, USA}
\affiliation{$^3$Department of Astronomy \& Astrophysics, The Pennsylvania State University, University Park, PA 16802, USA}
\affiliation{$^4$Institute for Gravitation and the Cosmos, The Pennsylvania State University, University Park, PA 16802, USA}
\affiliation{$^5$Center for Neutrino Physics, Department of Physics, Virginia Tech, Blacksburg, VA 24061, USA}
\affiliation{$^6$Kavli IPMU (WPI), UTIAS, The University of Tokyo, Kashiwa, Chiba 277-8583, Japan}
\affiliation{$^7$Key Laboratory of Particle Astrophysics and Experimental Physics Division and Computing Center, Institute of High Energy Physics, Chinese Academy of Sciences, Beijing 100049, China}
\affiliation{$^8$Department of Physics, Institute of Science Tokyo, 2-12-1 Ookayama, Meguro-ku, Tokyo 152-8551, Japan}

\date{submitted 1 June 2024; accepted 26 February 2026}

\begin{abstract}
We investigate the propagation of ultraheavy (UH) nuclei as ultrahigh-energy cosmic rays (UHECRs). We show that their energy loss lengths at $\lesssim300$~EeV are significantly longer than those of protons and intermediate-mass nuclei, and that the highest-energy cosmic rays with energies beyond $\sim100$~EeV, including the Amaterasu particle, may be UH-UHECRs. For the first time, we derive constraints on the contribution of UH-UHECR sources, and find that the current data are consistent with energy generation rate densities of UHECRs from collapsars and neutron star mergers. Our model predicts that the mean value of the depth of shower maximum is lower than that for iron nuclei beyond 100~EeV, which can be tested with future composition measurements, e.g., AugerPrime and the Global Cosmic Ray Observatory.  
In addition, the spectral tension between the Telescope Array (TA) and the Pierre Auger Observatory can be alleviated by considering the enhanced contribution of UHECRs -- including UH nuclei -- from a nearby transient. 
\end{abstract}

\maketitle

%
The origin of ultrahigh-energy cosmic rays (UHECRs) has been a long-standing mystery for over 50 years since the first detection of $\sim 100\rm~EeV$ cosmic rays (see reviews, Refs.~\citep{Hillas:1985is,Nagano:2000ve,Kotera:2011cp,AlvesBatista:2019tlv}).
The observed spectrum of UHECRs shows a hardening at the ankle around $4\rm~EeV$ and a cutoff at $\sim50\rm~EeV$~\cite{Abbasi:2007sv,Abraham:2008ru,Abraham:2010mj}.
There are possible discrepancies in the measured energy spectra~\cite{Ivanov:2017juh} between the Pierre Auger Observatory (Auger)~\cite{PierreAuger:2015eyc} and Telescope Array (TA)~\cite{Tokuno:2012mi,TelescopeArray:2012uws} observations.
Even though the differences at low energies may be explained by systematic effects in the energy scale, an excess remains in the TA data at the highest energies~\cite{PierreAuger:2023wti}. Recently, TA reported the detection of an extremely energetic UHECR event with the energy of $244 \pm 29 ({\rm stat.}) {}_{-76}^{+51} ({\rm syst.})\rm~EeV$, dubbed the ``Amaterasu'' particle~\citep{TelescopeArray:2023sbd}.

The composition of UHECRs is important to unveil their origin (see e.g., Ref.~\citep{PierreAuger:2022atd}). 
The depth of the cosmic-ray shower maximum, $X_{\rm max}$, is a measurable quantity used to infer the particle composition~\cite{Greisen:1960wc}.
The Auger data favor a mixed composition of UHECRs, and intermediate-mass (e.g., carbon and oxygen) and/or heavy (e.g., iron) nuclei make significant contributions beyond $10\rm~EeV$~\citep{PierreAuger:2010ymv, PierreAuger:2014sui,PierreAuger:2023xfc}.
In particular, the fraction of protons gradually decreases above the ankle, 
while intermediate-mass nuclei may become dominant at higher energies. 
The contribution of heavy nuclei seems negligible within the energy range of $10^{18.4}-10^{19.4}\rm~eV$~\cite{PierreAuger:2023xfc}, but these results are largely affected by hadronic interaction models. 
The distribution of $X_{\rm max}$ 
measured by TA is consistent with the Auger data~\cite{PierreAuger:2023yym}.

Not all UHECR source candidates can dominantly generate heavy nuclei as UHECRs. Production and acceleration of heavy nuclei has been considered in the context of collapsars~\cite{Murase:2008mr,Wang:2007xj,Murase:2010va,Metzger:2011xs,Liu:2011tv,Horiuchi:2012by,Zhang:2017moz,Zhang:2018agl,Boncioli:2018lrv,Bhattacharya:2021cjc} and compact binary mergers involving a neutron star~\cite{Takami:2013rza,Kyutoku:2016ckb,Kimura:2018ggg,Rodrigues:2018bjg,Murase:2018utn, Decoene:2019eux,Rossoni:2024ial}. The energy generation rate density of UHECRs at $10^{19.5}$~eV is $E Q_{E}^{19.5}\approx(0.2-2)\times 10^{43}\rm~erg~Mpc^{-3}~yr^{-1}$~(e.g., Refs.~\cite{Katz:2008xx,Jiang:2020arb}), which suffices for the energy budget of collapsars including gamma-ray bursts (GRBs) and hypernovae, as well as binary neutron star (BNS) mergers (see Table II of Ref.~\cite{Murase:2018utn}). Both the collapsar and BNS scenarios are consistent with the UHECR data if the energy fraction carried by cosmic rays is $\epsilon_{\rm CR}\sim0.1\%-10\%$. In addition, with the luminosity requirement~\cite{Blandford:1999hi} obtained from the Hillas condition~\citep{Hillas:1985is}, the maximum possible energy of the accelerated cosmic rays is estimated to be
\begin{align}
E_{A, \rm max}&\approx 9.8\times{10}^{20}~{\rm eV}~\left(\frac{Z}{40}\right) {\left(\frac{\epsilon_B}{0.01}\right)}^{1/2}  \nonumber \\
&\times  {\left(\frac{L}{{10}^{51}~{\rm erg/s}}\right)}^{1/2} {\left(\frac{\Gamma}{{10}^{2.5}}\right) }^{-1} \beta^{1/2}
\end{align}
where $Z$ is the nuclear charge, $\epsilon_B$ is the energy fraction of the magnetic field, $L$ is the total luminosity, $\Gamma$ is the Lorentz factor, and $\beta$ is the characteristic velocity normalized by the speed of light. 

Ultraheavy (UH) nuclei, which are defined as nuclei heavier than iron-group nuclei throughout this Letter, are believed to be synthesized due to the $r$ process occurring inside neutron-rich environments~\cite{Burbridge1957,Thielemann:2017acv,Horowitz:2018ndv,Kajino:2019abv,Cowan:2019pkx,Arcones:2022jer}. 
The sources of UH nuclei can be BNS and neutron-star--black-hole (NSBH) mergers (e.g., Refs.~\cite{Freiburghaus_1999,Wanajo:2014wha,Bartos:2019cec,Ekanger:2023mde,Fujibayashi:2022ftg}),
as well as collapsars including GRBs and magnetorotational supernovae (e.g., Refs.~\cite{Metzger:2011xs,Siegel:2018zxq,Barnes:2022dxv,Nishimura:2015nca, Yong:2021nkh,Bhattacharya:2021cjc,Ekanger:2022tia,Zha:2024fyo,Reichert:2024vyd}).
UH nuclei synthesized in magnetically dominated GRB outflows are discussed as UHECRs by Ref.~\cite{Metzger:2011xs}, and possible signatures of UHECR nuclei are investigated by Ref.~\cite{Murase:2010va}. UH nuclei have the advantage of being accelerated to energies beyond $100\rm~EeV$, compared to conventional light- and intermediate-mass group nuclei, which could provide an additional contribution to the highest-energy cosmic rays~\cite{Anchordoqui:1999aj}. 

In this Letter, we study the fate of UH-UHECRs during their intergalactic propagation and for the first time constrain their contributions to the highest-energy cosmic rays with energies exceeding $10^{20}$~eV. We also discuss the plausibility of the Amaterasu particle as an UH-UHECR event and its implications for future observations.

{\it Propagation of UH-UHECRs.---}
UH nuclei can be photodisintegrated and spalled into lighter nuclei due to their interactions with background target photons and matter, respectively. For the propagation of UHECRs from their source to Earth, interactions with the cosmic microwave background (CMB) and extragalactic background light (EBL) are dominant, and we utilize the public code {\sc CRPropa~3.2} to propagate UHECRs through intergalactic space~\cite{AlvesBatista:2022vem}. However, because {\sc CRPropa~3.2} does not provide a module for nuclei with mass numbers of $A >56$, we newly generate the photodisintegration cross section table of UH-UHECRs using the nuclear reaction network {\sc Talys~1.96}~\cite{Koning:2005ezu,refId0} as an extension to {\sc CRPropa~3.2}. In our photodisintegration reaction network, the maximum atomic number and mass number of nuclei are $Z=92$ and $A=238$, respectively, with a total of 2434 isotopes. In addition, the decay of unstable UH-UHECRs is implemented with the data table taken from {\sc NuDat~3}~\cite{NuDat3}.

\begin{figure}
    \centering
\includegraphics[width=\linewidth]{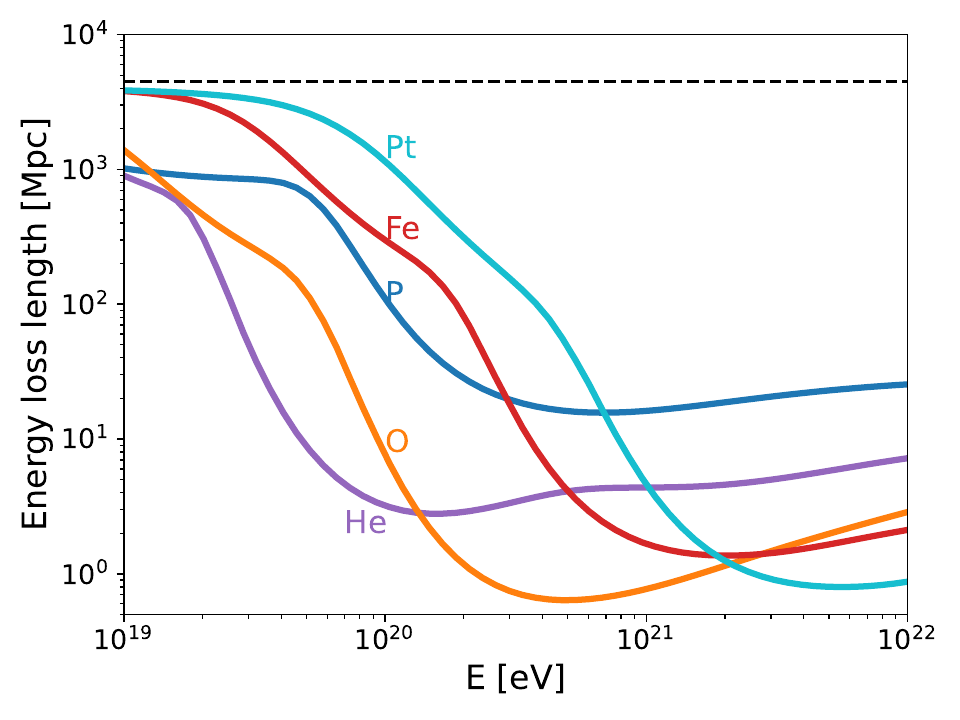}
    \caption{Total energy loss lengths for various nuclei: p, He, O, Fe, and Pt. The black dashed line is the energy loss length due to the adiabatic expansion of the universe. CMB and EBL~\cite{Gilmore:2011ks} are considered as target photons.}
    \label{fig:energy-loss-length-total}
\end{figure}

The dominant energy loss processes during the intergalactic propagation of UHECR nuclei are photodisintegration, photomeson production, Bethe-Heitler pair production, and adiabatic losses due to cosmic expansion. In Fig.~\ref{fig:energy-loss-length-total}, we show total energy loss lengths for light nuclei represented by proton (p) and helium (He), oxygen (O) as intermediate-mass nuclei, iron (Fe) as heavy nuclei, and platinum (Pt) as UH nuclei. See Supplementary Material~\footnotemark[1] for other representative UH nuclei. 
\footnotetext[1]{See Supplemental Material at \url{http://link.aps.org/supplemental/10.1103/221m-gvs3} for details of interactions of ultraheavy nuclei (e.g., photodisintegration cross sections, energy loss lengths, and propagation distance), fitting procedure, constraints from the best-fit models only with conventional nuclei, an enhanced contribution  from a nearby transient to explain the possible spectral tension between the Telescope Array and the Pierre Auger Observatory.}

The inverse of the total energy loss length is written as $\lambda_{\rm loss}^{-1} = \lambda_{\rm loss, phdis}^{-1} + \lambda_{\rm loss, phmes}^{-1} + \lambda_{\rm loss, BH}^{-1} + \lambda_{\rm loss, ad}^{-1}$, where $\lambda_{\rm loss, ad} = c/H_0\sim 4000\rm~Mpc$ is the adiabatic energy loss length. The photodisintegration energy loss length is $\lambda_{\rm loss}^{\rm phdis} \approx \left(n_{\rm CMB} \hat{\sigma}_{\rm phdis}\right)^{-1}\simeq 1.3 (A/195)^{-0.21}\rm~Mpc$, where $\hat{\sigma}_{\rm dis} = \sigma_{A\gamma} \kappa_{A\gamma}$ is the effective photodisintegration cross section and $\kappa_{A\gamma}\approx1 / A$ is the inelasticity at the giant dipole resonance (GDR)~\cite{Stecker:1969fw,Puget:1976nz}. The effective photodisintegration cross section for UH nuclei can be analytically approximated as $\sigma_{A\gamma} \approx \sigma_{\rm GDR} \Delta \bar{\varepsilon}_{\rm GDR} / \bar{\varepsilon}_{\rm GDR} \approx 120 (A/195)^{1.21}\,{\rm mb}$ for $A\gtrsim 4$, where $\sigma_{\rm GDR} \approx 4.3 \times 10^{-28} A^{1.35} \rm \ {cm}^2$ is the GDR cross section, $\Delta \bar{\varepsilon}_{\rm GDR} \approx 21.05 A^{-0.35} \rm~MeV$ is the width~\cite{EkangerSurvival}, and $\bar{\varepsilon}_{\rm GDR} \approx 42.65 A^{-0.21}\rm~MeV$ is the resonance energy in the nuclear rest frame~\cite{Karakula:1993he,Murase:2010gj}.
The typical resonance energy is $E_{A}^{\rm phdis} \approx 0.5 A m_p c^2 \bar{\varepsilon}_{\rm GDR} / \varepsilon_{t} \simeq 1.9\times 10^{21}~{\rm eV}~(A/195)^{0.79} (\varepsilon_{t} / 6.6\times 10^{-4}~{\rm eV})^{-1}$, where $\varepsilon_t$ is the target photon energy. 
EBL is more relevant to the photodisintegration of UH nuclei with $\sim 10^{20}\rm~eV$.
Energy losses due to the Bethe-Heitler pair production process are also important for UH-UHECRs because of their large atomic numbers. The energy loss length is estimated to be $\lambda_{\rm loss}^{\rm BH} \approx \left(n_{\rm CMB} \hat{\sigma}_{\rm BH}\right)^{-1} \simeq 32\ (Z/78)^{-2} (A/195)\rm~Mpc$, where $\hat{\sigma}_{\rm BH}\sim8\times 10^{-31} (Z^2/A)\rm~cm^2$ is the effective cross section~\cite{Murase:2010va,Murase:2018iyl}, $E_{A}^{\rm BH} \approx 0.5 A m_p c^2 \bar{\varepsilon}_{\rm BH} / \varepsilon_{\rm CMB} \simeq 1.4\times 10^{21}~{\rm eV}~(A/195)(\varepsilon_{t}/6.6\times 10^{-4}\rm~eV)^{-1}$ is the typical energy, and $\bar{\varepsilon}_{\rm BH} \sim 10\rm~MeV$. 
Although the photodisintegration process is expected to be dominant, which is indeed the case at $E_A\sim{\rm a~few}\times10^{21}$~eV, we find that the Bethe-Heitler pair production process is the most important at $E_A \sim (1 - 5) \times 10^{20}\rm~eV$ (see Fig.~S1 in Supplemental Material~\footnotemark[1]).
The photomeson production process is irrelevant when considering the propagation of UH-UHECRs due to the high-energy threshold at $\sim 4\times 10^{22}~{\rm~eV}~(A/195)(\varepsilon_{t}/6.6\times 10^{-4}\rm~eV)^{-1}$.

As seen in Fig~\ref{fig:energy-loss-length-total}, UH-UHECRs can travel longer distances than the GZK (Greisen-Zatsepin-Kuzmin) distance for UHECR protons~\cite{Greisen:1966jv,Zatsepin:1966jv} and the energy loss length of iron-group nuclei~\cite{Stecker:1969fw,Puget:1976nz,Stecker:1997rs, Stecker:1998ib, Metzger:2011xs}. 
Thanks to their large mass numbers, it takes longer for them to be completely disintegrated, by which rarer sources can contribute or UH-UHECRs could appear beyond the cutoff in spectra of UHECR protons and conventional nuclei. To calculate the UHECR spectrum and composition on Earth, we take into account the change in the composition of UH-UHECRs during propagation, as the UH nuclei become progressively lighter with the ejection of one or more nucleons (see Supplemental Material for details). 

{\it UH-UHECR spectrum and composition.---}
Using the available Auger and TA data for the energy spectrum and composition of UHECRs, we examine possible contributions of the sources of UH-UHECRs. For the demonstration, we consider two representative composition models of conventional nuclei, depending on whether there are heavy iron-group nuclei or not:
(i) Model A: conventional nuclei (p, He, O, Si) + UH nuclei (Se, Te, Pt); (ii) Model B:  conventional nuclei (p, He, O, Si, Fe) + UH nuclei (Se, Te, Pt).
The origin of intermediate-mass nuclei can be associated with material from the inner core of massive stars collapsing into black holes~\citep{Zhang:2017hom,Zhang:2018agl}. Iron-group nuclei are naturally expected in the supernova or hypernova ejecta~\citep{Murase:2008mr,Wang:2007xj,Liu:2011tv,Horiuchi:2012by,Zhang:2018agl,Bhattacharya:2021cjc} or the surface of neutron stars~\cite{Fang:2012rx,Fang:2013vla}.  Alternatively, UHECR nuclei can be provided by the reacceleration of Galactic cosmic rays~\cite{Caprioli:2015zka,Kimura:2017ubz}. 
We treat UH-UHECRs as an additional component, assuming that they are produced by either the same or different populations. We consider Selenium (Se), Tellurium (Te), and Platinum (Pt) to represent the first, second, and third peaks of UH nuclei synthesized via the $r$ process.  

\begin{figure*}
    \centering
    \includegraphics[width=\textwidth]{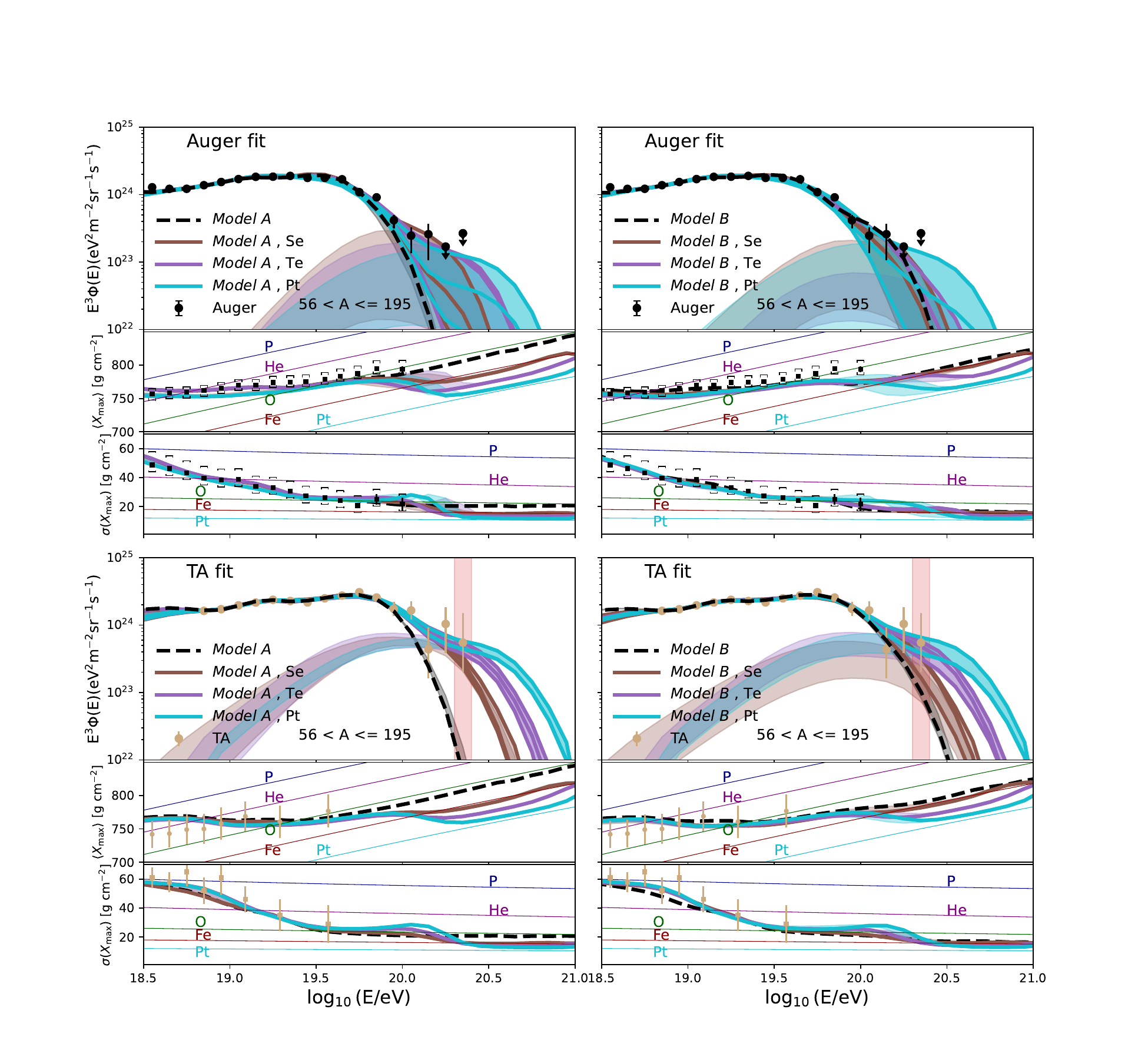}
    \caption{Energy spectra and the first/second moments of $X_{\rm max}$ distribution are shown considering both conventional and UH nuclei. Auger data are obtained from Refs.~\cite{PierreAuger:2020qqz,PierreAuger:2024flk}, while TA data are from Refs.~\cite{TelescopeArray:2024tbi,TelescopeArray:2018xyi}. Note that the data point in the red vertical band corresponds to the Amaterasu particle~\cite{TelescopeArray:2023sbd}.}
    \label{fig:Diffuse_AugerTA_bestfit}
\end{figure*}

The injection spectrum of UHECRs escaping from the sources is assumed to be a power law with an exponential cutoff,
\begin{equation}\label{eq:spectrum}
    \frac{d\dot{\mathcal{N}_A}}{d\mathcal{R}} = f_A \dot{\mathcal{N}_0} \left(\frac{\mathcal{R}}{ \mathcal{R}_0}\right)^{-s_{\rm CR}} {\rm exp}\left(-\frac{\mathcal{R}}{\mathcal{R}_{\rm max}}\right),
\end{equation}
where $\mathcal{R}=E/Z$ is the particle rigidity, $E$ is particle energy, and $Z$ is nuclear charge, $\mathcal{R}_0$ is a reference rigidity, $f_A$ is the number fraction of nuclei, $s_{\rm CR}$ is the spectral index, and $\dot{\mathcal{N}}_0$ is a normalization constant.
To reduce the number of free parameters, we assume all of the accelerated nuclei have the same rigidity, $\mathcal{R}_{\rm max}$, and the spectral index, $s_{\rm CR}$.
We also assume no redshift evolution for the sources, which is sufficient for our purpose to constrain the contribution of UH-UHECRs. 
Large-scale magnetic fields in structured regions of the Universe could further modify the observed spectrum and composition, whereby cosmic rays with low rigidities cannot arrive on Earth, also known as the magnetic horizon effect~\cite{Lemoine:2004uw,Berezinsky:2007kz}. However, this effect is important only if the average source separation is $d_s \gtrsim 20 \sqrt{l_c / 0.1\rm~Mpc}\rm~Mpc$, where $l_c$ is the coherence length~\cite{PierreAuger:2024hlp}. Because transient sources like collapsars and BNS mergers may lead to sufficient source densities~\cite{Murase:2008sa,Takami:2011nn}, it is reasonable to adopt the one-dimensional propagation scenario assuming UHECR sources have an average distance less than $20\rm~Mpc$. 
In this work, we adopt the hadronic interaction model {\sc Epos-LHC}~\citep{Pierog:2013ria, PierreAuger:2022atd}.

To constrain the energy generation rate density of UH-UHECRs, $Q_{\rm UH-UHECR}\equiv \int_{E_{\rm min}}^{E_{\rm max}} d\mathcal{R} E \frac{d\dot{N_A}}{d\mathcal{R}}$ with $E_{\rm min} = 10^{18}\rm~eV$ and $E_{\rm max} = 10^{21}\rm~eV$ (cf. Ref.~\cite{Murase:2011cy}),  
we perform a combined fit with both conventional and UH nuclei components. For simplicity, we only consider the combination of conventional nuclei with one of the UH nuclei, among Se, Te, and Pt, for both \textit{Model A} and \textit{Model B}, which allows us to explore the effect of mass and charge of UH nuclei on the derived fit.
For the TA data, we show the best-fit values of the energy generation rate density of UH-UHECRs, where 
the uncertainty is estimated under the condition that the total chi-square value $\chi_{\rm tot}^2 \leqslant \chi_{\rm tot, min}^2 + 1$.
For the Auger data, we use $\chi_{\rm tot}^2 \leqslant \chi_{\rm tot, min}^2 + 10$ to estimate the uncertainty range because of the much smaller statistical errors in the measured flux and composition.
Our results are summarized in Table~\ref{tab:ultraheavynuclei}. 

In Fig.~\ref{fig:Diffuse_AugerTA_bestfit}, we provide the results of our fitting for the energy spectrum and composition of UHECRs measured by Auger~\cite{PierreAuger:2020qqz,Yushkov:2020nhr} (see SM for details). With \textit{Model A} and \textit{Model B}, we see that the energy generation rate density of the three UH nuclear species is constrained to be 
\begin{equation}
Q_{\rm UH-UHECR}^{\rm Auger} \lesssim (0.1 - 15) \times 10^{42}\rm~erg~Mpc^{-3}~yr^{-1}.
\end{equation}
The best-fit values derived from \textit{Model B} can be about one order of magnitude smaller than those from \textit{Model A},
because the presence of heavy iron-group nuclei gives tighter constraints on the fraction of UH-UHECRs. We note that the values have significant uncertainties, and more stringent constraints on the energy budget of UH-UHECRs are obtained when we use the best-fit models only with conventional nuclei (see SM for details). 

\begin{table}[]
\def\arraystretch{1.25}
    \centering
    \begin{tabular}{lcc}
    \hline
    Nuclei & $Q_{\rm UH-UHECR}^{\rm Auger}$ & $Q_{\rm UH-UHECR}^{\rm TA}$\\
    & $[\rm 10^{42}\rm~erg~Mpc^{-3}~yr^{-1}]$ &$[\rm 10^{43}\rm~erg~Mpc^{-3}~yr^{-1}]$ \\
\hline
    \textit{Model A} \\
    Se   &  $2.6_{-1.8}^{+12.6}$ &  $4.1_{-0.9}^{+1.5}$\\
    Te   &  $5.4_{-5.3}^{+0.7}$ &  $2.9_{-1.0}^{+0.8}$\\
    Pt   &  $1.5_{-0.7}^{+3.1}$&  $2.0_{-0.0}^{+0.9}$\\
\hline
    \textit{Model B} \\
    Se   &  $1.0_{-0.0}^{+10.8}$&  $5.1_{-3.7}^{+0.3}$\\
    Te   &  $2.7_{-2.2}^{+3.0}$ &  $2.9_{-1.1}^{+0.9}$\\
    Pt   &  $1.1_{-1.0}^{+4.4}$ &  $1.7_{-0.0}^{+1.5}$\\ \hline
    \end{tabular}
    \caption{Energy generation rate densities of UH-UHECRs, allowed by the two composition models used in this work.}
    \label{tab:ultraheavynuclei}
\end{table}

The results for the TA data are shown in Fig.~\ref{fig:Diffuse_AugerTA_bestfit}, where we find that the energy generation rate densities of UH-UHECRs are constrained to be 
\begin{equation}
Q_{\rm UH-UHECR}^{\rm TA} \lesssim (1.4 - 5.6) \times 10^{43}\rm~erg~Mpc^{-3}~yr^{-1},
\end{equation}
for both composition models, which is about 3 times larger than that derived based on the Auger data for both composition models. The upper limits from the Auger data are tighter than those from the TA data. This is because the TA spectrum includes the data point from the Amaterasu particle and the TA composition is subject to larger statistical errors. There may be a significant hemispheric variability in the physical properties of the sources, and it is also important to understand differences between the TA and PAO spectra (see Fig.~S5 of Supplemental Material).

We use the Akaike information criterion (AIC) for model selection~\cite{1100705, Burnham2004}.
In particular, we find that including UH-UHECRs as a second population can provide better fits to the observed energy spectrum and composition measured by TA, with a rescaled AIC value of $\Delta_i  = {\rm AIC}_i - {\rm AIC}_{\rm min} \lesssim 2$. This is compared to fits with only conventional nuclei, where $\Delta_i = 4.3$ for Model A and $\Delta_i = 3.3$ for Model B. We use ${\rm AIC}_{\rm min} = 31.0$ for Model A + Pt as the baseline for comparison. 
However, including UH-UHECRs does not improve the fits to Auger data, where $\Delta_i \gtrsim 8$. In this case, we use ${\rm AIC}_{\rm min} = 70.5$ for Model A as the baseline for comparison.
While UH-UHECRs help us better explain the highest-energy UHECR data, more conservatively, our results are also regarded as constraints on their possible contribution. We stress that not only the spectrum but also the composition data at the highest energies are critical for deriving our constraints on UH-UHECR energetics. As indicated in Fig.~\ref{fig:Diffuse_AugerTA_bestfit}, a significant contribution of UH-UHECRs at the highest energies predicts the decrease of $\langle X_\text{max} \rangle$ and $\sigma(X_\text{max})$ above $\sim100$~EeV. With future UHECR measurements by AugerPrime~\cite{Castellina:2019irv} and GCOS~\cite{GCOS:2023yyr}, examining the saturation of the composition data at the iron line can be an interesting test for the BNS and collapsar scenarios for UH-UHECRs.  

\begin{figure}
    \centering
 \includegraphics[width=\linewidth]{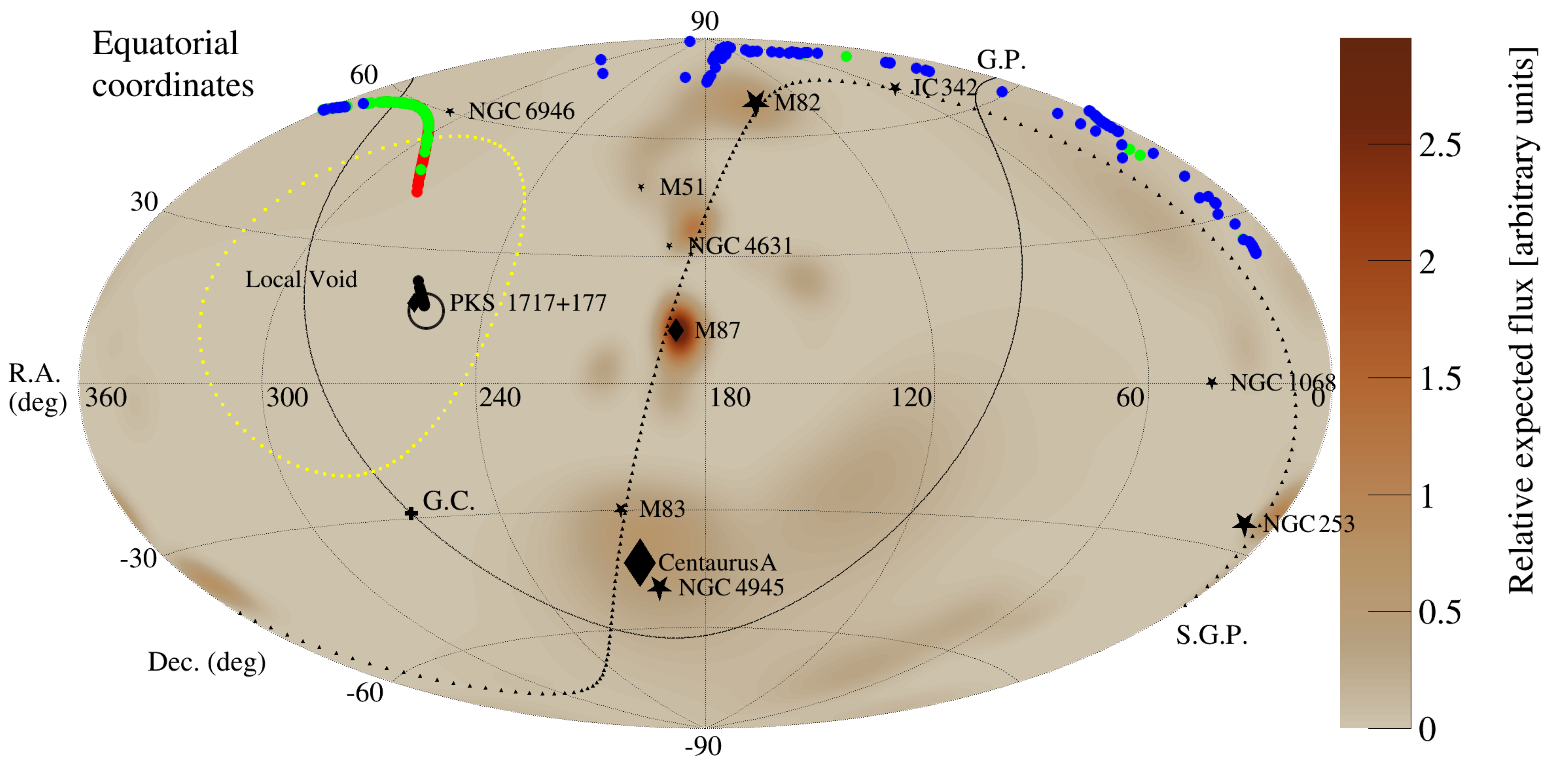}
    \caption{Skymap of backtracked particles with mean energy $E = 244\rm~EeV$ and variation $E=70\rm~EeV$ for p ($Z = 1$, black), Fe ($Z = 26$, red), Zr ($Z = 40$, green) and Pt ($Z = 78$, blue) in equatorial coordinates. For each nuclear species, we inject 100 particles. The arrival direction of the Amaterasu particle is (R.A., Dec.) = ($255.9\pm 0.6^\circ$, $16.1\pm 0.5^\circ$) in equatorial coordinates, indicated as a black circle. The supergalactic plane (S.G.P.) is shown as black dotted curves, and the Galactic plane (G.P.) is shown as black solid curves. The color bar represents the expected relative flux from sources in the local large-scale structure.
    This figure is generated with the code provided by Ref.~\cite{TelescopeArray:2023sbd}.}
    \label{fig:skymap}
\end{figure}

It is interesting to consider the Amaterasu particle as a possible UH-UHECR event. Even though the primary photon is excluded at the 99.985\% confidence level, it is difficult to distinguish whether it is a proton or heavy nucleus~\citep{TelescopeArray:2023sbd}. 
Indeed, as shown in Fig.~\ref{fig:Diffuse_AugerTA_bestfit}, the Amaterasu particle may be explained as a UH-UHECR event. To explore this possibility in more detail, we examine the backtracked direction of the Amaterasu particle for different nuclear species, as shown in Fig.~\ref{fig:skymap}, where we adopt the Galactic magnetic field model provided in Ref.~\cite{Jansson:2012rt}. For light or even iron nuclei, the direction of the Amaterasu particle lies in the local void region (yellow dotted curve in Fig.~\ref{fig:skymap})~\cite{TelescopeArray:2023sbd, Unger:2023hnu,Kuznetsov:2023jfw}. If it is a UH nucleus, the source may exist outside the local void or even near the supergalactic plane thanks to the larger atomic number. 

Our derived quantity, the UH-UHECR energy generation rate density, is crucial for testing source models. The current data suggest $Q_{\rm UH-UHECR}\sim 10^{43}\rm~erg~Mpc^{-3}~yr^{-1}$ as the best-fit values, being consistent with both collapsars and BNS/NSBH mergers~\cite{Murase:2018utn}. The cosmic-ray luminosity density of BNS mergers is $\sim10^{43.5}\rm~erg~Mpc^{-3}~yr^{-1}$, which is consistent with the kinetic energy $\mathcal{E}_{\rm ej}\approx(1/2) M_{\rm ej} c^2\beta_{\rm ej}^2 \sim 2\times10^{51}~{\rm erg}~(M_{\rm ej}/0.05~M_{\odot}){(\beta_{\rm ej}/0.2)}^2$, the rate density $\rho\sim300\rm~Gpc^{-3}~yr^{-1}$~\cite{KAGRA:2021duu,Sarin:2022cmu}, and $\epsilon_{\rm CR}\sim5$\%. Short GRBs have isotropic-equivalent energies of ${\mathcal E}_{\gamma}^{\rm iso}\sim10^{51}-10^{52}~{\rm erg}$ and rate densities of $\rho\sim10~{\rm Gpc}^{-3}~{\rm yr}^{-1}$~\cite{Wanderman:2014eza,Escorial:2022nvp}, inferring that the gamma-ray luminosity density is $\sim{10}^{43}-{10}^{44}~{\rm erg}~{\rm Mpc}^{-3}~{\rm yr}^{-1}$. Thus, it is possible for UH nuclei from BNS mergers and short GRBs to contribute to UHECRs above $\sim10^{20}$~eV. The energy budget requirement for UH-UHECRs is less demanding, so collapsars such as long GRBs~\cite{Milgrom:1995um,Waxman:1995dg} are also viable. 
Future UHECR data above $\sim100$~EeV as well as better theoretical understanding of $\epsilon_{\rm CR}$ for UH nuclei will enable us to critically test these UH-UHECR models.

{\it Summary and discussions.---}
We presented the first detailed study on the propagation of UH-UHECRs, and derived general constraints on their contribution to the observed UHECR flux. Thanks to their energy loss lengths at $\lesssim10^{21}$~eV, which are longer than those of protons and intermediate-mass nuclei, UH-UHECRs may significantly contribute to the highest-energy cosmic rays beyond $\sim10^{20}$~eV, including the Amaterasu event. The allowed energy generation rate densities are consistent with those of collapsars and compact binary mergers~\cite{Murase:2018utn}. 

Establishing the existence of UH nuclei at the highest energies may indicate that UHECRs are produced by transients rather than steady sources such as active galactic nuclei. UH nuclei as seeds for UHECRs can be synthesized in both BNS mergers and collapsars~\cite{Metzger:2011xs,Bhattacharya:2021cjc,Ekanger:2022tia,Ekanger:2023mde}. Note that the UHECR acceleration should occur in outflows or ejecta~\footnotemark[2]
\footnotetext[2]{The transrelativistic SN model~\cite{Zhang:2018agl} predicts that the maximal rigidity is similar among sources, which is consistent with the recent analysis~\cite{Ehlert:2022jmy}. Note that prompt gamma rays are produced in a different region.} 
rather than at external forward shocks (e.g., Refs.~\cite{Takami:2013rza,Kimura:2018ggg,Rodrigues:2018bjg}). Promising acceleration sites include external reverse shocks~\cite{Murase:2008mr,Zhang:2018agl} or internal dissipation~\cite{Murase:2008mr,Zhang:2017hom,Boncioli:2018lrv}. More detailed implications for the sources, including the nucleus survival problem, will be discussed in future work~\cite{EkangerSurvival}.

The UH nuclear origin of UHECRs could be tested with future measurements of the composition at the highest energies, as indicated in Fig.~\ref{fig:Diffuse_AugerTA_bestfit}. This behavior is in contrast to a high-luminosity GRB model that predicts a lighter composition at the highest energies~\cite{Zhang:2017moz} and a reacceleration model that changes toward an iron-group composition~\cite{Kimura:2018ggg}. 
Multimessenger observations with neutrinos and gamma rays would also be useful. This scenario also requires the survival of nuclei inside the sources, in which TeV gamma rays are likely to escape from the sources~\cite{Murase:2008mr}. Neutrinos may come from the beta decay of nuclei~\cite{Murase:2010gj}, while the gamma-ray signal from nuclear deexcitation and electromagnetic cascades induced by the Bethe-Heitler process could be interesting targets for future gamma-ray detectors such as the Cherenkov Telescope Array~\cite{Murase:2010va}. 

We also point out that the discrepancy between the Auger and TA spectra can be alleviated by considering an enhanced contribution from a nearby transient. As described in SM (see Fig.~S5), the TA spectrum may be explained by the best-fit model to the Auger data (\textit{Model A}) with an additional contribution of UHECRs—including UH nuclei—from a nearby low-luminosity GRB located at a distance $5\rm~Mpc$ from Earth. In this model, the conventional nuclei could come from the inner core of the progenitor star~\cite{Zhang:2017moz}, while the UH nuclei could be synthesized in the jet. A BNS/NSBH merger is another possible nearby transient that can alleviate the tension between the Auger and TA spectra.

{\it Note added. ---} 
Recently, a related work~\cite{Farrar:2024zsm} appeared regarding the BNS/NSBH merger scenario.
Our Letter is independent and focuses on the general aspects of UH-UHECRs that can be produced by not only BNS mergers but also collapsars.
Also, regarding the TA-Auger tension, the effect of a nearby source was independently studied by Refs.~\cite{Bartos:2025xjs,Korochkin:2025zvz}.

\acknowledgements
We thank Toshihiro Fujii, Kunihito Ioka, Michael Unger, and Xilu Wang for useful information and discussions. 
This work was supported by NSF Grants Nos.~AST-1908689 (K.M. and M.B.) and AST-1908960 (N.E., M.B. and S.H.). We also acknowledge NSF Grants Nos.~AST-2108466 (K.M.), AST-2108467 (K.M.), AST-2308021 (K.M.), and PHY-2209420 (S.H.), and U.S.~DOE Office of Science award number DE-SC0020262 (S.H.), and KAKENHI Nos.~20H01901 (K.M. and B.T.Z.), 20H05852 (K.M. and B.T.Z.), 22K03630 (S.H.), and 23H04899 (S.H.). 
B. T. Z. acknowledges the Yukawa Institute for Theoretical Physics, Kyoto University, where this work was completed before he moved to his current affiliation.
M.B. acknowledges support from the Eberly Postdoctoral Fellowship at the Pennsylvania State University. This work was supported by World Premier International Research Center Initiative (WPI Initiative), MEXT, Japan. 
K.M. also acknowledges Mamoru Yanagisawa for his generous donation and continuous support.

\bibliography{bzhang}

\clearpage
\setcounter{figure}{0}
\renewcommand{\thefigure}{S\arabic{figure}}

\section{SUPPLEMENTAL MATERIAL}\label{sup}

\subsection{Interactions of UH Nuclei}
\begin{figure}[h]
    \centering
\includegraphics[width=\linewidth]{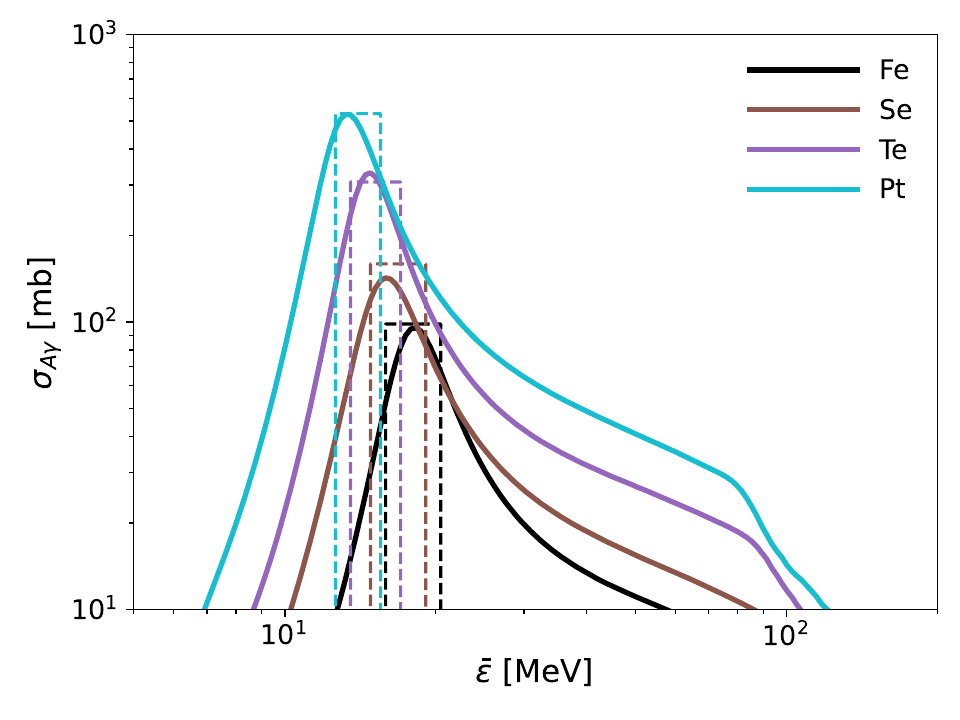}
    \caption{Photodisintegration cross sections for UH nuclei as a function of photon energy in the nuclear rest frame. Solid curves are from the output of {\sc Talys~1.96}, while the dashed boxes are estimated using the rectangular approximation~\cite{EkangerSurvival}.}
    \label{fig:cross section}
\end{figure}

\begin{figure}
\centering
\includegraphics[width=\linewidth]{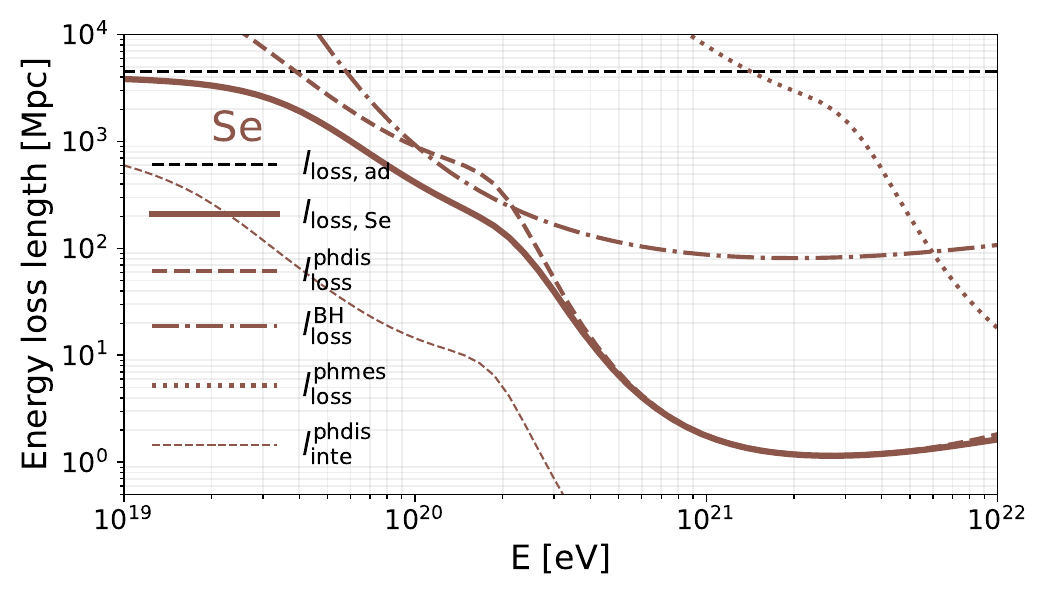}
\includegraphics[width=\linewidth]{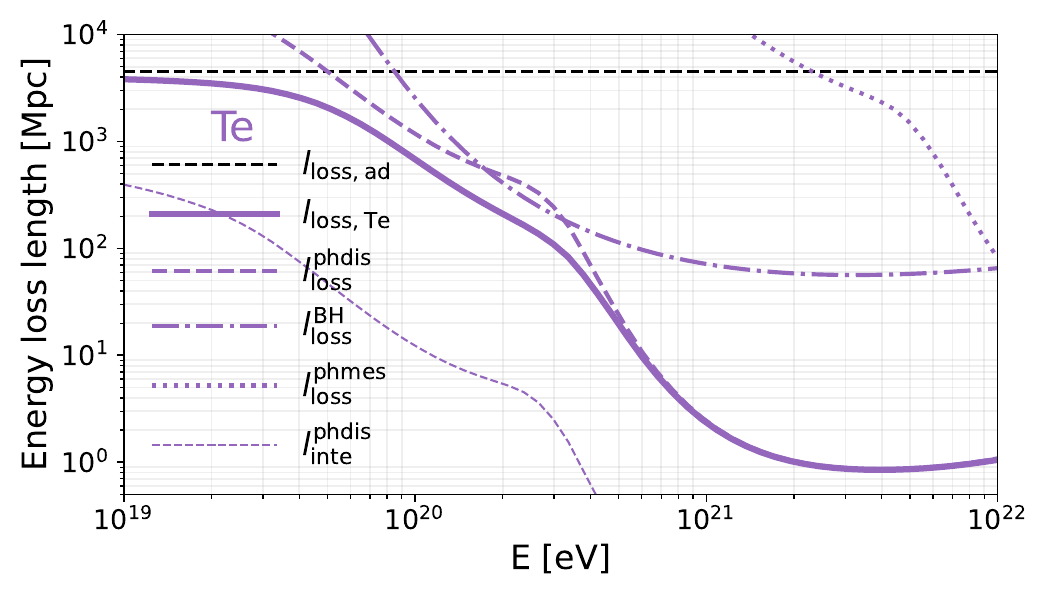}
\includegraphics[width=\linewidth]{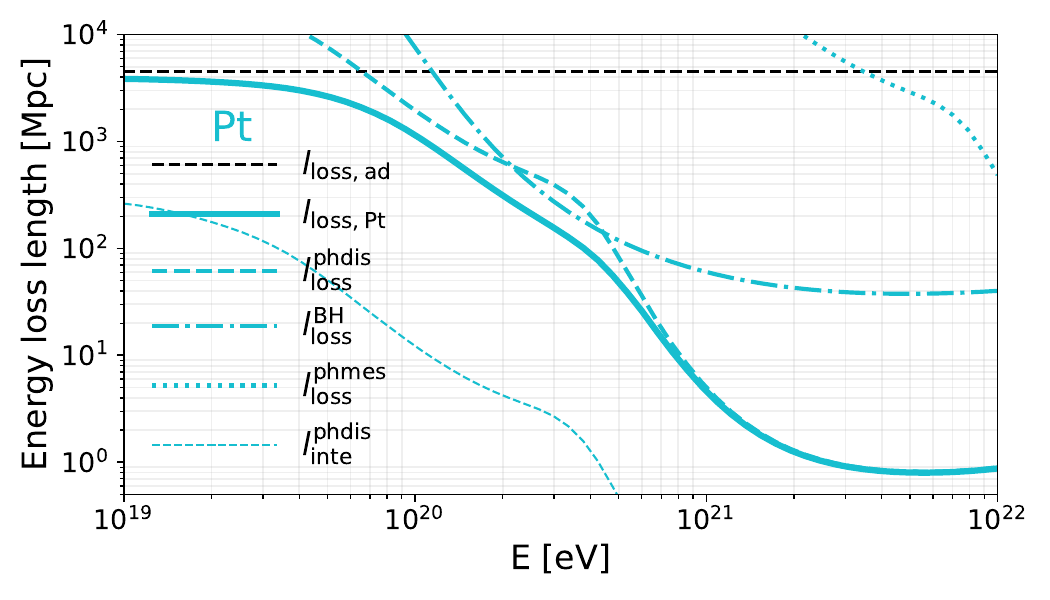}
    \caption{Energy loss lengths for photodisintegration (thick dashed lines), photomeson production (dotted lines), Bethe-Heitler pair production (dotted-dashed lines), and adiabatic expansion (black-dashed line) of UH nuclei of Se, Te, and Pt, as a function of nuclear energy. We also indicate interaction lengths for the photodisintegration process (thin dashed lines).}
    \label{fig:energy-loss-length}
\end{figure}

Photodisintegration cross sections for UH nuclei are based on the theoretical models of {\sc Talys~1.96/2.0}~\cite{talys2023} and are shown in Fig.~\ref{fig:cross section}. 
TALYS also uses experimental data (for example, cross sections), and supplements this with a theoretical model when experimental data is unavailable~\cite{talys2023}.
The rectangular approximation to the {\sc Talys} models is used to analytically evaluate energy loss lengths~\cite{EkangerSurvival}.
We also show the photodisintegration cross section of iron nuclei for comparison. We see a reasonably good match between the analytical estimate and the numerical results. In Fig.~\ref{fig:energy-loss-length}, we show the energy loss length for three typical UH nuclei, namely Selenium (Se), Tellurium (Te), and Platinum (Pt), respectively. 
Nuclear reaction models could affect the energy loss length at a level of 20\% for Silicon~\cite{Kido:2022ads}. The effect of the nuclear reaction models for UH nuclei should be further studied. The extension of \textsc{CRPropa 3.2} with UH nuclei developed for this work is available at \href{https://github.com/btheodorezhang/CRPropa3-Source.git}{github.com/btheodorezhang/CRPropa3-Source.git}.

\begin{figure}
    \centering
    \includegraphics[width=\linewidth]{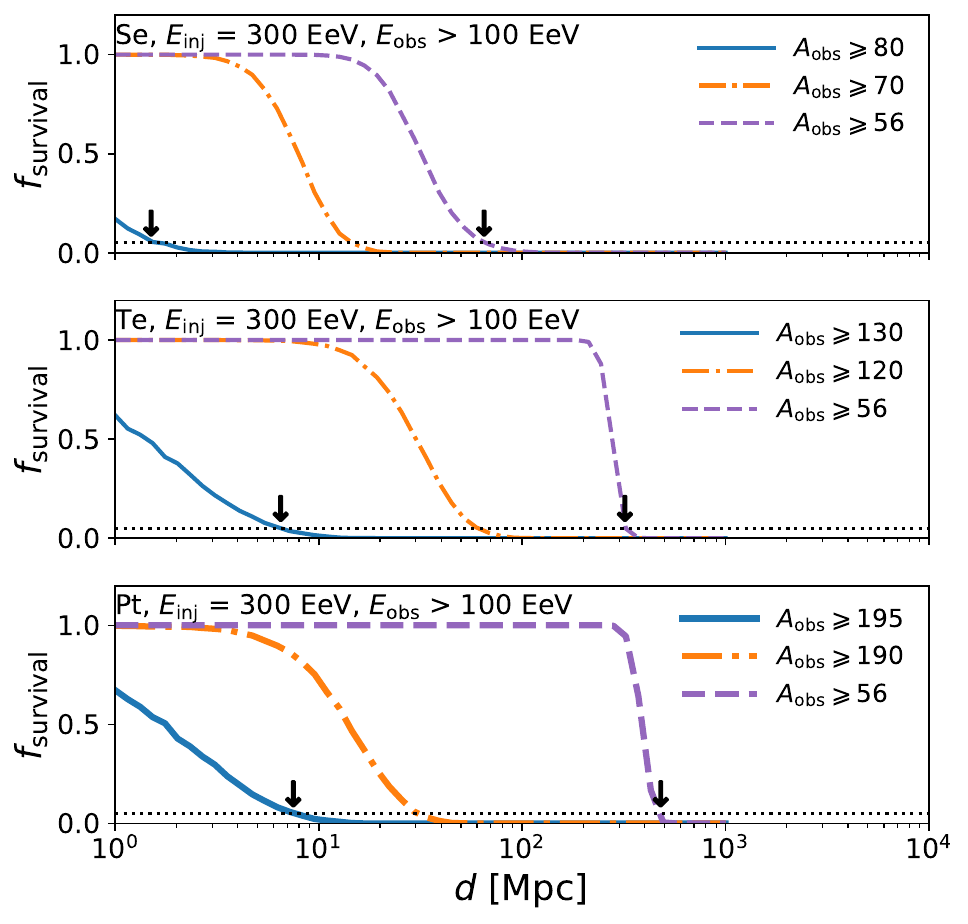}
    \caption{Propagation distance, $d_{95\%}$, corresponding to $f_{\rm survival} = 5\%$, is shown for Se, Te, and Pt. The black-dashed lines correspond to a value of 0.05 indicating that 95\% of the nuclei are lost during propagation.}
    \label{fig:mfp}
\end{figure}

To study the change in the composition of UH-UHECRs during propagation, we estimate the fraction of surviving nuclei $f_{\rm survival} = N_{\rm obs} (E > E_{\rm obs}, A > A_{\rm obs})/N_{\rm inj} (E_{\rm inj})$ numerically with {\sc CRPropa 3}~\citep[e.g.,][]{Globus:2022qcr}, 
where $N_{\rm inj}$ is the number of nuclei injected at a given distance, $N_{\rm obs}$ is the observed number of particles, $A_{\rm obs}$ and $E_{\rm obs}$ are the observed nuclei mass number and energy, respectively. 
The propagation distance $d_{95\%}$ is defined by  $f_{\rm survival} = 5\%$, which implies that $95\%$ of the nuclei are lost during propagation.
In Fig.~\ref{fig:mfp}, we show the propagation distance $d_{95\%}$ for three typical UH nuclei species.
For UH nuclei with primary energy $E_{\rm inj}=300\rm~EeV$, the propagation distance is $d_{95\%} \sim 1.5 \rm~Mpc$ for Se ($A_{\rm obs} \geqslant 80$), $d_{95\%} \sim 6.5 \rm~Mpc$ for Te ($A_{\rm obs} \geqslant 130$) and $d_{95\%} \sim 7.5 \rm~Mpc$ for Pt ($A_{\rm obs} \geqslant 195$). 
The propagation distance increases to $d_{95\%}\approx 65\rm~Mpc$ for Se  ($A_{\rm obs} \geqslant 56$), $d_{95\%}\approx 320\rm~Mpc$ for Te  ($A_{\rm obs} \geqslant 56$) and $d_{95\%}\approx 480\rm~Mpc$ for Pt ($A_{\rm obs} \geqslant 56$).

\begin{table*}[]
    \centering
    \begin{tabular}{lcccccccc} \hline
     & \textit{Model A} & \textit{Model A} & \textit{Model A} & \textit{Model A} & \textit{Model B} & \textit{Model B} & \textit{Model B} & \textit{Model B} \\
    & & (+ Se) &  (+ Te) & (+ Pt) & & (+ Se) & (+ Te) & (+ Pt)  \\
    \hline
    \textbf{Auger} & & & & & & \\
$Q_{44}$                 &  5.94     &   5.94   &  6.01    & 4.99    & 5.87    &  5.05   &  4.82   &  4.99 \\
$\mathcal{R}_{\rm max}$ &  18.1     &   18.1   &  18.1    & 18.2    & 18.1    &  18.3   &  18.3   &  18.3 \\
$s_{\rm CR}$            &  -2.2     &   -2.2   &  -2.2    & -0.7    & -2.1    &  -0.4   &  -0.3   &  -0.4 \\
$f_{\rm P}$             &  0.23\%   &   0.80\% &  2.04\%  & 1.97\%  & 1.52\%  &  1.74\% &  8.71\% &  0.84\% \\
$f_{\rm He}$            &  58.25\%   &   52.72\% &  57.77\%  & 52.41\%  & 53.77\%  &  35.37\% &  14.20\% &  27.30\% \\
$f_{\rm O}$             &  37.16\%   &   42.70\% &  36.85\%  & 35.04\%  & 42.92\%  &  52.75\% &  69.00\% &  62.29\% \\
$f_{\rm Si}$            &  4.35\%   &   3.71\% &  3.27\%  & 10.55\%  & 0.91\%  &  9.45\% &  6.43\% &  8.83\% \\
$f_{\rm Fe}$            &  -   &   - &  -  & -  & 0.87\%  &  0.65\% &  1.65\% &  0.72\% \\
$f_{\rm Se}$            &  -   &   0.07\% &  -  & - & -  &   0.04\% &  - &  - \\
$f_{\rm Te}$            &  -   &   - &  0.08\%  & -  & -  &  - &  0.01\% &  - \\
$f_{\rm Pt}$            &  -   &   - &  -  & 0.02\%  & -  &  - &  - &  0.02\% \\
$\chi_{\rm min}^2$      &  56.1   & 62.4 &  61.7  & 71.3  & 57.5 &  67.0 &  68.2 &  65.9 \\
$\chi_{\rm min}^2/{\rm d.o.f.} (\rm d.o.f.)$  &  1.66(36)  & 1.78(35) &  1.76(35)  & 2.04(35)  & 1.64(35) &  1.97(34) &  2.01(34) &  1.94(34) \\
AIC           &  70.5  & 79.6 &  78.9  & 88.6  & 74.8 &  87.3 &  88.6 &  86.3 \\
$\Delta_i$         &  0.0  & 9.1 &  8.4  & 18.1  & 4.3 &  16.8 &  18.1 &  15.8 \\
    \hline
    \textbf{TA} & & & & & & \\
$Q_{44}$                 &  7.83     &   7.57   &  6.14    & 6.29    & 7.74    &  6.43   &  6.07   &  5.80 \\
$\mathcal{R}_{\rm max}$ &  18.4     &   18.4   &  18.6    & 18.6    & 18.4    &  18.6   &  18.6   &  18.7 \\
$s_{\rm CR}$            &  -0.6     &   -0.4   &  0.4    & 0.3    & -0.5    &  0.4   &  0.4   &  0.5 \\
$f_{\rm P}$             &  3.25\%   &   11.27\% &  54.67\%  & 41.67\%  & 1.57\%  &  31.88\% &  55.40\% &  15.25\% \\
$f_{\rm He}$            &  76.50\%   &   74.32\% &  22.78\%  & 25.75\%  & 80.82\%  &  51.45\% &  16.75\% &  5.75\% \\
$f_{\rm O}$             &  0.43\%   &   0.93\% &  0.44\%  & 0.36\%  & 1.18\%  &  0.26\% &  3.63\% &  2.54\% \\
$f_{\rm Si}$            &  19.82\%   &   12.89\% &  21.74\%  & 31.98\%  & 15.40\%  &  15.21\% &  23.50\% &  75.22\% \\
$f_{\rm Fe}$            &  -   &   - &  -  & -  & 1.03\%  &  0.30\% &  0.29\% &  0.80\% \\
$f_{\rm Se}$            &  -   &   0.59\% &  -  & - & -  &   0.90\% &  - &  - \\
$f_{\rm Te}$            &  -   &   - &  0.38\%  & -  & -  &  - &  0.44\% &  - \\
$f_{\rm Pt}$            &  -   &   - &  -  & 0.23\%  & -  &  - &  - &  0.45\% \\
$\chi_{\rm min}^2$      &  18.9   & 12.2 &  11.0  & 10.8  & 14.1 &  13.6 &  12.0 &  12.0 \\
$\chi_{\rm min}^2/{\rm d.o.f.} (\rm d.o.f.)$  &  0.94(20)  & 0.64(19) &  0.58(19)  & 0.57(19)  & 0.74(19) &  0.75(18) &  0.66(18) &  0.67(18) \\
AIC           &  35.3  & 32.4 &  31.2  & 31.0  & 34.3 &  38.1 &  36.4 &  36.4 \\
$\Delta_i$         &  4.3  & 1.4 &  0.2  & 0.0  & 3.3 &  7.1 &  5.4 &  5.4 \\
       \hline
    \end{tabular}
    \caption{Best-fit parameters, where $Q_{\rm 44} \equiv Q_{\rm UHECR} / 10^{44}\rm~erg~Mpc^{-3}~yr^{-1}$ and $Q_{\rm UHECR}$ is the total energy generation rate density including UH-UHECRs.}
    \label{tab:parameters-bestfit}
\end{table*}

\subsection{Fitting Procedure}
Here, we describe the details of the fitting procedure for the observed UHECR spectra and composition.
The observed flux is estimated with the following formula~\cite[e.g.,][]{Zhang:2017hom},
\begin{equation}
    \Phi_A = \sum_{A^\prime} \Phi_{A A^\prime} = \sum_{A^\prime} \frac{c}{4\pi}\frac{dn_{A A^\prime}}{dE},
\end{equation}
and
\begin{eqnarray}
dn_{A A^\prime}(E) &=& \int_{z_{\rm min}}^{z_{\rm max}} dz  \left| \frac{dt}{dz}\right| \xi(z) \nonumber \\ &\times& \int_{E^\prime_{\rm min}}^{E^\prime_{\rm max}} dE^\prime \frac{d\dot N_{A^\prime}}{dE^\prime}\eta_{A A^\prime}(E, E^\prime,z),
\end{eqnarray}
where $\eta_{A A^\prime}(E, E^\prime,z)$ is a factor representing the number of secondary UHECRs with mass number A and energy E that are generated from parent nuclei with the mass number $A^\prime$ and $E^\prime$~\cite{Khan:2004nd,Unger:2015laa,Zhang:2017hom}, $z$ is redshift, $\xi(z)$ is the redshift evolution factor, and $dt/dz = -1/[H_0 (1 + z) (\sqrt{\Omega_\Lambda + \Omega_k(1 + z)^2 + \Omega_m(1 + z)^3})]$.
We assume a continuous source distribution with a minimum source distance of $d_{\rm min} = 1.5\rm~Mpc$ and a maximum source distance of $d_{\rm max} = 3854\rm~Mpc$, corresponding to redshifts, $z_{\rm min} = 0.0003$ and $z_{\rm max} = 4.5$, respectively. 
Note that the distances, $d_{\rm min}$ and $d_{\rm max}$, should be treated as light travel distances.
If this distance is larger, the cutoff of the conventional component is lower for a given parameter set, and we find that the UH-UHECR component is more favored. While shorter distances can compete with the effect of UH-UHECR nuclei, there is no Milky-Way-like galaxy within 1.5 Mpc in the northern sky, so it is reasonable and conservative to assume 1.5 Mpc as the minimum distance.
The minimum injection energy is $E^\prime_{\rm min} = 10^{18}\rm~eV$, and the maximum injection energy is $E^\prime_{\rm max} = 10^{21}\rm~eV$.
The cosmological parameters adopted in this work are $H_0 = 67.3 \ \rm km \ s^{-1} \ Mpc^{-1}$, $\Omega_m = 0.315$, $\Omega_\Lambda = 0.685$, and $\Omega_k = 0$~\cite{Agashe:2014kda}.

We adopt the chi-square method to fit the observed energy spectrum and composition~\cite{Jiang:2020arb}, and use
\begin{eqnarray}
\chi^2_\text{spec}=&&\sum_i \frac{(f \Phi^\text{model}(\hat{E}_i; s_{\rm CR}, \mathcal{R}_\text{max})-\Phi(E_i))^2}{\Delta(\Phi)_i^2}\nonumber\\ &&+\left(\frac{\delta_E}{\sigma_E}\right)^2,
\end{eqnarray}
where $f$ is a free normalization parameter, $\Phi^\text{model}$ is the simulation results, $\hat{E}_i\equiv (1+\delta_E)E_i$, $\Delta(\Phi)$ is the statistic uncertainty, and $\sigma_E$ is the systematic uncertainty on the measured energy scale by Auger ($\sigma_E \sim 14\%$)~\cite{PierreAuger:2020qqz} and TA ($\sigma_E \sim 21\%$)~\cite{TelescopeArray:2023bdy}. 
In this work, the systematic uncertainty on the energy scale is not considered by setting $\delta_E = 0$. We note that the variation of the observed flux $E^3 \Phi$ has an uncertainty of $\sim 30\%$ ($\sim 50\%$) for Auger (TA)~\cite{PierreAuger:2022atd}. Correspondingly, our constraints on the energy generation rate densities of UH-UHECRs in Table~I and Table~\ref{tab:ultraheavynuclei-fixed} would be subject to this systematics.

Similarly, we estimate the $\chi^2$ value when fitting to the composition. For the first and second moments of $X_{\rm max}$ distribution, $\langle X_\text{max} \rangle$ and $\sigma(X_\text{max})$, we have
\begin{align}
\chi^2_{\langle X_\text{max} \rangle} &= \sum_i \frac{(\langle X_\text{max}^\text{model}\rangle(\hat{E}_i; s_{\rm CR}, \mathcal{R}_\text{max})-\langle X_\text{max}\rangle(E_i))^2}{\Delta(X_{\rm max})_i^2},
\end{align}

\begin{align}
\chi^2_{\sigma(X_\text{max})} &= \sum_i \frac{(\sigma(X_\text{max})(\hat{E}_i; s_{\rm CR}, \mathcal{R}_\text{max})-\sigma(X_\text{max})(E_i))^2}{\Delta(\sigma(X_\text{max}))_i^2}, 
\end{align}
where $\Delta(X_{\rm max})$ and $\Delta(\sigma(X_\text{max}))$ are the uncertainties including both statistical and systematic uncertainties~\cite[e.g.,][]{Fang:2017zjf}.
The values of $\langle X_\text{max} \rangle$ and $\sigma(X_\text{max})$ are calculated based on the generalized Gumbel distribution function~\cite{DeDomenico:2013wwa}, with the updated parameters for {\sc Epos-LHC} from Table 5 of ~\cite{PierreAuger:2022atd}. We assume that the parameters could be extended to UH nuclei, which is reasonable in the superposition model. The range of the value of $X_\text{max}$ we considered is from 100 to 1500.
The total $\chi^2$ value is 
\begin{equation}
    \chi^2 = \chi^2_\text{spec}+\chi^2_{\langle X_\text{max} \rangle}+\chi^2_{\sigma(X_\text{max})},
\end{equation}
and the best-fit results are obtained when $\chi^2$ reaches its minimum value $\chi^2_{\rm min}$.
In this work, we fit the observed spectrum and composition with energies above $10^{18.75}\rm~eV$ for Auger data. There are $16$ data points for the spectrum and $26$ data points for the first and second moments of the $X_{\rm max}$ distribution.
For TA data, we consider energies above $10^{18.85}\rm~eV$, with $16$ data points on the spectrum and $10$ data points for the first and second moments of the $X_{\rm max}$ distribution.
The degree of freedom ($\rm d.o.f.$) is defined as the total number of data points minus the number of fitting parameters.
We treat the maximum particle energy ($\mathcal{R}_{\rm max}$), spectral index ($s_{\rm CR}$), energy generation rate density ($Q_{\rm UHECR}$), and number fraction of nuclei ($f_A$), as the free parameters.
Note that the fractions, satisfying $\sum f_A = 1$, are generated as the direction cosines in $k$ dimensions using the same method described in Ref.~\cite{PierreAuger:2016use}.
In Table~\ref{tab:parameters-bestfit}, we show the best-fit parameters used in Fig.~2.

\begin{figure*}
    \centering
    \includegraphics[width=\textwidth]{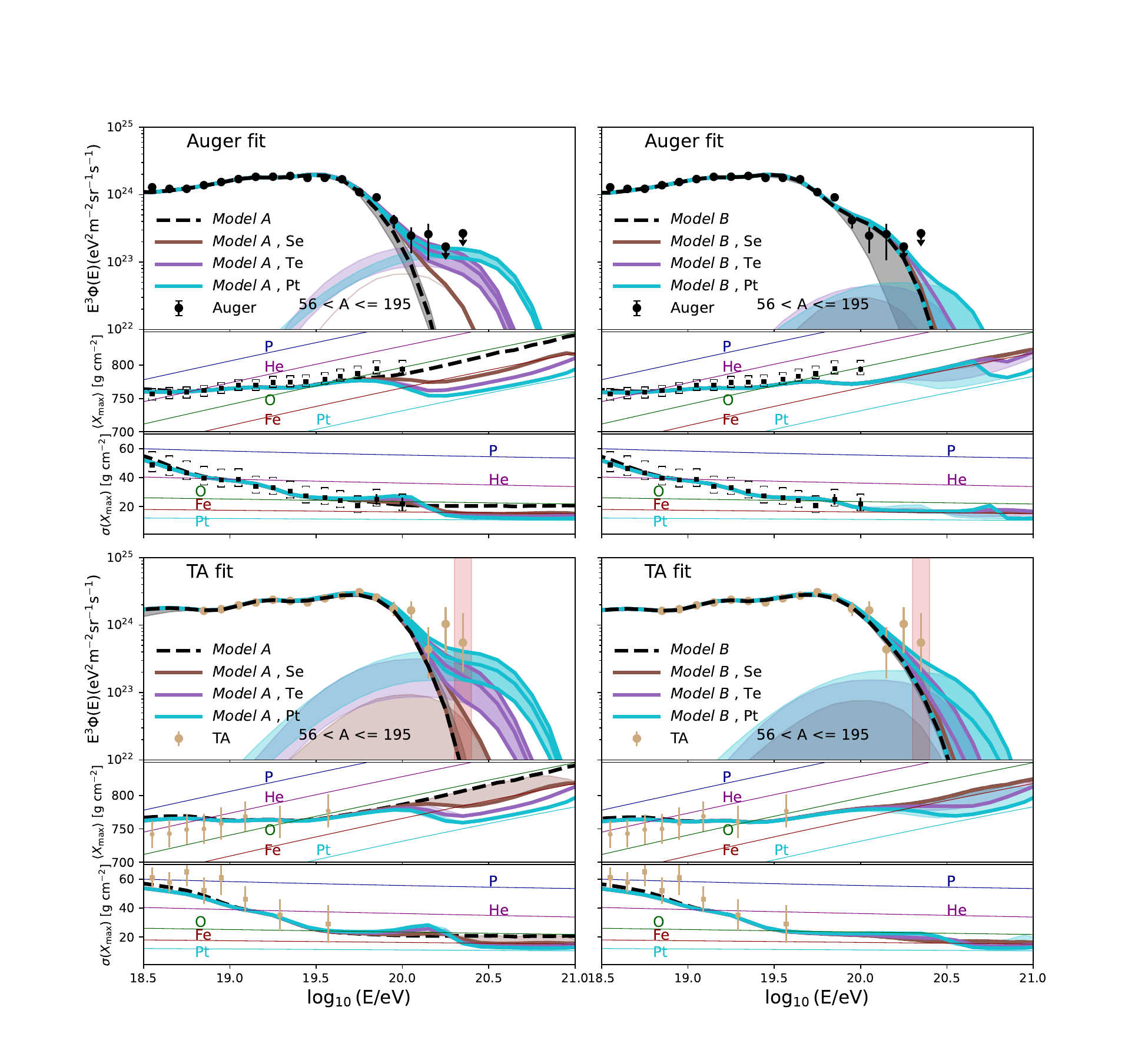}
    \caption{Energy spectra and the first/second moments of $X_{\rm max}$ distribution considering both conventional and UH nuclei. 
    Auger data are obtained from Refs.~\cite{PierreAuger:2020qqz,PierreAuger:2024flk}, while TA data are from Refs.~\cite{TelescopeArray:2024tbi,TelescopeArray:2018xyi}.
    }
    \label{fig:Diffuse_AugerTA_fixed}
\end{figure*}

\subsection{Constraints from the Best-Fit Models Only with Conventional Nuclei}

We constrain the energy generation rate density of UH-UHECRs, $Q_{\rm UH-UHECR}$, based on the best-fit model using only conventional nuclei.
Specifically, we begin with the best-fit results considering only conventional nuclei, summarized in Table~\ref{tab:parameters-bestfit}. Then, we inject UH-UHECRs as an additional source population, but with the same rigidity and spectral index.
Our results are shown in Fig.~\ref{fig:Diffuse_AugerTA_fixed}. We can see that the obtained results provide more stringent constraints compared to the combined fit method, as shown in Table~\ref{tab:ultraheavynuclei-fixed}.
With \textit{Model A}, the energy generation rate densities of these three UH nuclei are constrained to a narrow range, with $Q_{\rm UH-UHECR}^{\rm Auger} \sim (4 - 8) \times 10^{42}\rm~erg~Mpc^{-3}~yr^{-1}$.
However, we only obtain the upper limits on the UH-UHECR energy budget when adopting \textit{Model B}, $Q_{\rm UH-UHECR}^{\rm Auger}\lesssim (1.2-1.4)\times 10^{42}\rm~erg~Mpc^{-3}~yr^{-1}$.
The reason is that including heavy iron-group nuclei provides a better fit for the observed energy spectra, with $\chi_{\rm tot, min}/{\rm d.o.f.}=64/{\rm d.o.f.}$ for conventional nuclei in \textit{Model B} and $\chi_{\rm tot, min}/{\rm d.o.f.}=72/{\rm d.o.f.}$ in \textit{Model A}, where $\rm d.o.f.$ is the degree of freedom.
With TA data, we obtain $Q_{\rm UH-UHECR}^{\rm TA} \sim (1 - 3) \times 10^{43}\rm~erg~Mpc^{-3}~yr^{-1}$ for \textit{Model A} and $Q_{\rm UH-UHECR}^{\rm TA} \lesssim (0.5 - 1.3) \times 10^{43}\rm~erg~Mpc^{-3}~yr^{-1}$ for \textit{Model B}, which is about 3 times larger than that derived based on the Auger data.

\begin{table}[]
    \centering
    \begin{tabular}{lcc}
    \hline
    Nuclei & $Q_{\rm UH-UHECR}^{\rm Auger}$ & $Q_{\rm UH-UHECR}^{\rm TA}$\\
    & $[\rm 10^{42}~erg~Mpc^{-3}~yr^{-1}]$ &$[\rm 10^{43}~erg~Mpc^{-3}~yr^{-1}]$ \\
\hline
\textit{Model A}, fixed \\
    Se  & {$3.9_{-1.7}^{+1.4}$} &{$5.3_{-5.2}^{+5.2}$}\\
    Te   & {$4.4_{-1.5}^{+1.2}$} &{$0.9_{-0.5}^{+0.5}$}\\
    Pt    & {$5.0_{-1.32}^{+0.0}$} & {$1.1_{-0.5}^{+0.5}$}\\
\hline
    \textit{Model B}, fixed \\
    Se   & {$\lesssim 1.5$} &{$\lesssim (0.37 - 0.5)$}\\
    Te   & {$\lesssim 1.5$}& {$\lesssim (0.19 - 0.64)$}\\
    Pt   & {$\lesssim 1.6$} & {$\lesssim (0.32 - 0.82)$}\\ \hline
    \end{tabular}
    \caption{Energy generation rate densities of UH-UHECRs constrained from the best-fit models
only with conventional nuclei.}
    \label{tab:ultraheavynuclei-fixed}
\end{table}

\subsection{Enhanced Contribution from A Nearby Transient}\label{appendix:transient}
\begin{figure}
    \centering
    \includegraphics[width=0.45\textwidth]{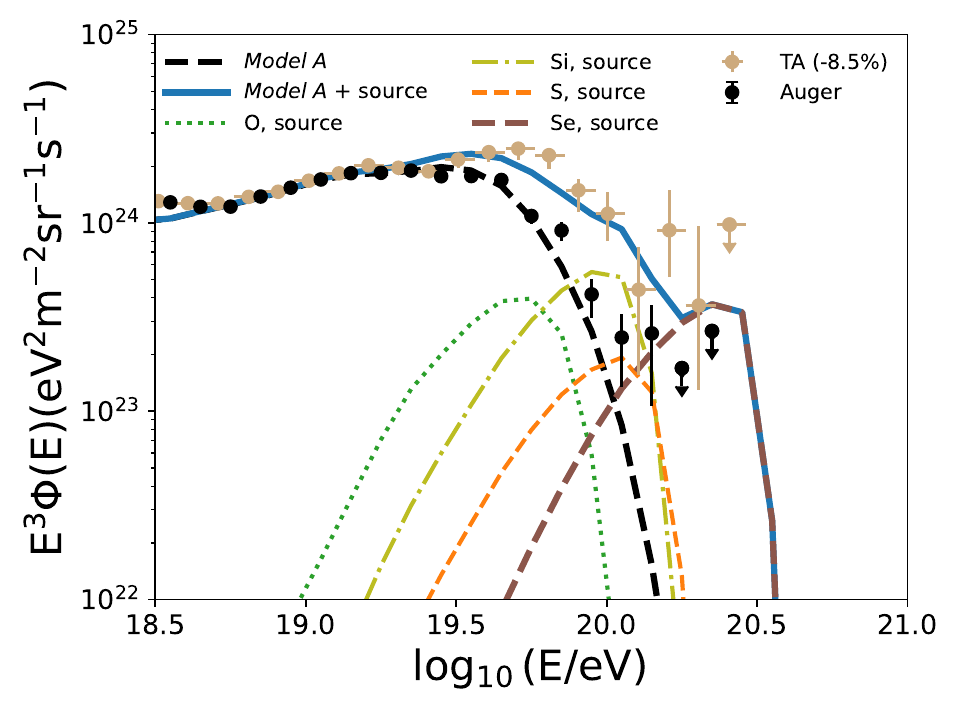}
    \caption{Illustrative example for explaining the TA spectrum with the best-fit model to the Auger data ($Model \ A$) with an additional contribution of UHECRs including UH nuclei from collapsars.}
    \label{fig:Auger-TA-source}
\end{figure}

A nearby transient may give a special contribution to the observed UHECR flux, and we explore the possibility of simultaneously explaining both the Auger and TA spectra. A specific example is shown in Fig.~\ref{fig:Auger-TA-source}, where the flux per steradian is calculated by dividing the total flux by the detector's field of view, $\Delta \Omega \sim 2\pi$. The observed TA spectral excess may be affected by the systematic uncertainty on the energy scale, and we decrease the TA energy scale by $8.5\%$ to match the Auger spectra in the low-energy range~\cite{Plotko:2022urd, PierreAuger:2023wti}.
Based on the best-fit model to the Auger data, we assume there are some nearby transient sources in the northern sky, mainly coming from the supergalactic plane, accounting for the spectral excess observed by TA.
For demonstrative purposes, we assume a nearby low-luminosity GRB located at $d = 5\rm~Mpc$ from Earth, and adopt the 16TJ model shown in Table I of Ref.~\cite{Zhang:2017moz} as the default composition model of nuclei, $f_{\rm O} : f_{\rm Si} : f_{S} = 0.52 : 0.37 : 0.11$, and the total injection UHECR luminosity above $10^{18}$~eV is $\mathcal{L}_{\rm UHECR} \simeq 0.7\times 10^{41}\rm~erg~s^{-1}$.
Additionally, UH nuclei could be synthesized in the relativistic jet of GRBs~\cite{Metzger:2011xs,Bhattacharya:2021cjc}, where the most abundant UH nuclei can be the first-peak $r$ process elements such as Se, and the corresponding injection luminosity is $\mathcal{L}_{\rm UH-UHECR} \simeq 0.7\times 10^{40}\rm~erg~s^{-1}$.
The typical delay time of UHECRs due to the extragalactic magnetic field can be $\tau_d^{\rm EG} \sim 1.7~{\rm~yr}~(Z/34)^2 E_{A, 20.5}^{-2} B_{\rm EG, -11.5}^2 (l_c/1{\rm~Mpc}) (d/{5\rm~Mpc})$, where $B_{\rm EG}$ is the magnetic field strength that can be very weak in the void region, and $d$ is source distance (e.g., Refs.~\cite{Murase:2008mr,Murase:2009ah}).
The Galactic magnetic field causes an inevitable time delay, $\tau_d^{\rm Gal} \sim 200\rm~yr$ for a particle with rigidity $5\rm~EV$~\cite{Murase:2008sa,Takami:2011nn,Marafico:2024qgh}.
Thus, the total required energy of UHECRs emitted by a transient source is estimated to be $\mathcal{E}_{\rm UHECR} \sim (\mathcal{L}_{\rm UHECR} + \mathcal{L}_{\rm UH-UHECR}) (\tau_d^{\rm EG} + \tau_d^{\rm Gal}) \simeq 4.5\times 10^{50}\rm~erg$, which can be consistent with the cosmic-ray energy budgets of collapsars~\cite{Murase:2008mr,Zhang:2017moz,Zhang:2018agl}.
Note that even in this specific case our constraints on the energy generation rate density of UH-UHECRs still hold, and the results with the Auger data are applicable. 

\end{document}